\documentclass[prb,preprint,aps,showpacs]{revtex4}

\usepackage{graphicx}
\def\be{\begin{equation}}
\def\ee{\end{equation}}
\def\bdm{\begin{displaymath}}
\def\edm{\end{displaymath}}
\def\ba{\begin{array}}
\def\ea{\end{array}}
\def\bea{\begin{eqnarray}}
\def\eea{\end{eqnarray}}

\begin{document}

\title{Spin current and spin accumulation near a Josephson junction between a singlet and triplet superconductor}

\author{Chi-Ken~Lu and Sungkit~Yip}

\affiliation{Institute of Physics, Academia Sinica, Nankang, Taipei
115, Taiwan}

\date{\today }

\begin{abstract}

We consider a Josephson junction with an arbitrary transmission
coefficient $\mathcal{D}$ between a singlet and a triplet
superconductor with the latter order parameter characterized by a
d-vector of the form ($k_x\hat{y}-k_y\hat{x}$). Various quantities
such as the tunnelling current, spin accumulation, and spin
current are calculated via the quasiclassical Green's functions.
We also present a symmetry argument on the existence of these
quantities and their dependencies on the phase difference across
the junction. A physical picture is also given in terms of the
Andreev states near the junction.

\end{abstract}


\maketitle

\section{Introduction}

Recently, there has been much interest in manipulating the spin
degree of freedom of electrons in condensed matter systems.
Phenomena such as spin current, spin Hall effect, spin
accumulation and magneto-electric effects have received a lot of
attention.\cite{review}   These phenomena have been discussed in a
variety of systems, including metals, semiconductors, and even
insulators. In this paper, we discuss the spin current and spin
accumulation near a Josephson junction.  We shall in particular
consider a junction between an $s$-wave superconductor and a pure
triplet superconductor with the latter in the state where the
"d-vector" specifying the spin structure of the Cooper pairs be
given by $\hat{d}=k_x\hat{y}-k_y\hat{x}$.

We are interested in this $k_x\hat{y}-k_y\hat{x}$ state for a
number of reasons. This state corresponds to the one-dimensional
representation $A_{2u}$ in a crystal with tetragonal $D_{4h}$
symmetry,\cite{group} and hence is one of the simplest example of
a triplet state. This state is also believed to be a limiting case
for the order parameter of the non-centrosymmetric
superconductor\cite{Bauer-r} CePt$_{x}$Pd$_{3-x}$Si. There, due to
the absence of inversion symmetry in the normal state, the order
parameter is believed to be a mixture of $s$-wave and the $p$-wave
$A_{2u}$ order parameter. [The state $k_x\hat{y}-k_y\hat{x}$ is
the limiting case (perhaps for small Pd concentration $3-x$) where
the s-wave admixture is small.] Similar mixing of superconducting
order parameter of different parity is also expected in compounds
such as CeRhSi$_3$,\cite{Ce113} and in superconductivity found at
oxide interface.\cite{Reyren} Spin current generated near the
surface of this superconducting state with vacuum have been
discussed recently by two groups,\cite{Vorontsov,Tanaka} with and
without the mixing of the s-wave order parameter due to the
absence of inversion symmetry.

We generalized these considerations to the case where this
superconductor is in contact with an $s$-wave superconductor in
the form of a Josephson junction with arbitrary transmission
coefficient $\mathcal{D}$, but for simplicity we shall not include
any broken inversion symmetry effects in the normal state, hence
the bulk superconductors are assumed to be pure singlet and pure
triplet, respectively. Clearly, in the infinitely high barrier
limit, our results would just be a special case of
Ref.\cite{Vorontsov,Tanaka}.

For general transmission however, one expects a proximity effect
so that near the interface, the system acquires properties of a
superconductor with mixed singlet and triplet order parameters,
similar to the case which arises in non-centrosymmetric
superconductors,\cite{Bauer-r,Ce113,Reyren} even though our bulk
superconductors are each purely singlet and triplet. Effects that
are normally not allowed can now appear due to the lowering of
symmetries, similar to the electro-magneto effects discussed
recently for bulk non-centrosymmetric superconductors.\cite{e-m}
There, in particular, a supercurrent can generate a spin
polarization in a perpendicular direction. Here, we shall
investigate how the spin current (and the spin accumulation)
depends on (and hence can be manipulated by) the phase difference
between these two superconductors.

Our investigation is interesting in another point of view. The
state $k_x\hat{y}-k_y\hat{x}$ has two counter-propagating edge
states of opposite spins near a surface (see below), in direct
analogy with the quantum spin Hall state often discussed in the
current literature.\cite{QSH,QSH2}   Our investigations here then
is analogous to considering an interface between an ordinary
"insulator" (our $s$-wave superconductor) and a "quantum spin-Hall
insulator" (our $k_x\hat{y}-k_y\hat{x}$ superconductor).
Discussions on this and other related triplet superconductors from
this point of view can also be found in Ref.\cite{helicalSC,Sato}.

A recent paper\cite{Sengupta} also studies the spin accumulation
near a Josephson junction between a pure singlet and pure triplet
superconductor.  In that paper, only the very special $p$-wave
state where $\hat{d}$ is independent of the momentum direction
$\hat{k}$ was considered. Spin accumulation was shown to exist
near the junction, with the spin direction along $\hat d$. The
authors suggested the detection of this spin accumulation as a
method of identifying triplet superconductors. However, the
constant $\hat d$ vector is a very special case.  A general
triplet superconductor is expected to have $\hat{k}$ dependent $d$
vectors.\cite{group} For these more general cases, it is then
unclear if spin accumulation would exist, and in which direction
the net spin lies. We would like to provide a general
consideration using this ($k_x\hat{y}-k_y\hat{x}$) state as an
illustrative example.

Our paper is organized as follows.  We begin with a symmetry
argument in Sec.\ref{symmetry}.  We then present our calculations
with the quasiclassical method in Sec.\ref{main}.  The subsections
provides our results, first for the special cases of perfect and
small transmissions, then the more general case with arbitrary
$\mathcal{D}$. We summarize in Sec.\ref{summary}. We employ a
generalization of the "exploding and decaying trick", which we
explain in Appendix\ \ref{trick}.

%Appendix \ref{symmetry} provides the general symmetry arguments.

\section{The junction geometry and symmetry considerations}\label{symmetry}

\begin{figure}
\input{epsf}
\epsfxsize=6in \epsfysize=4in \epsfbox{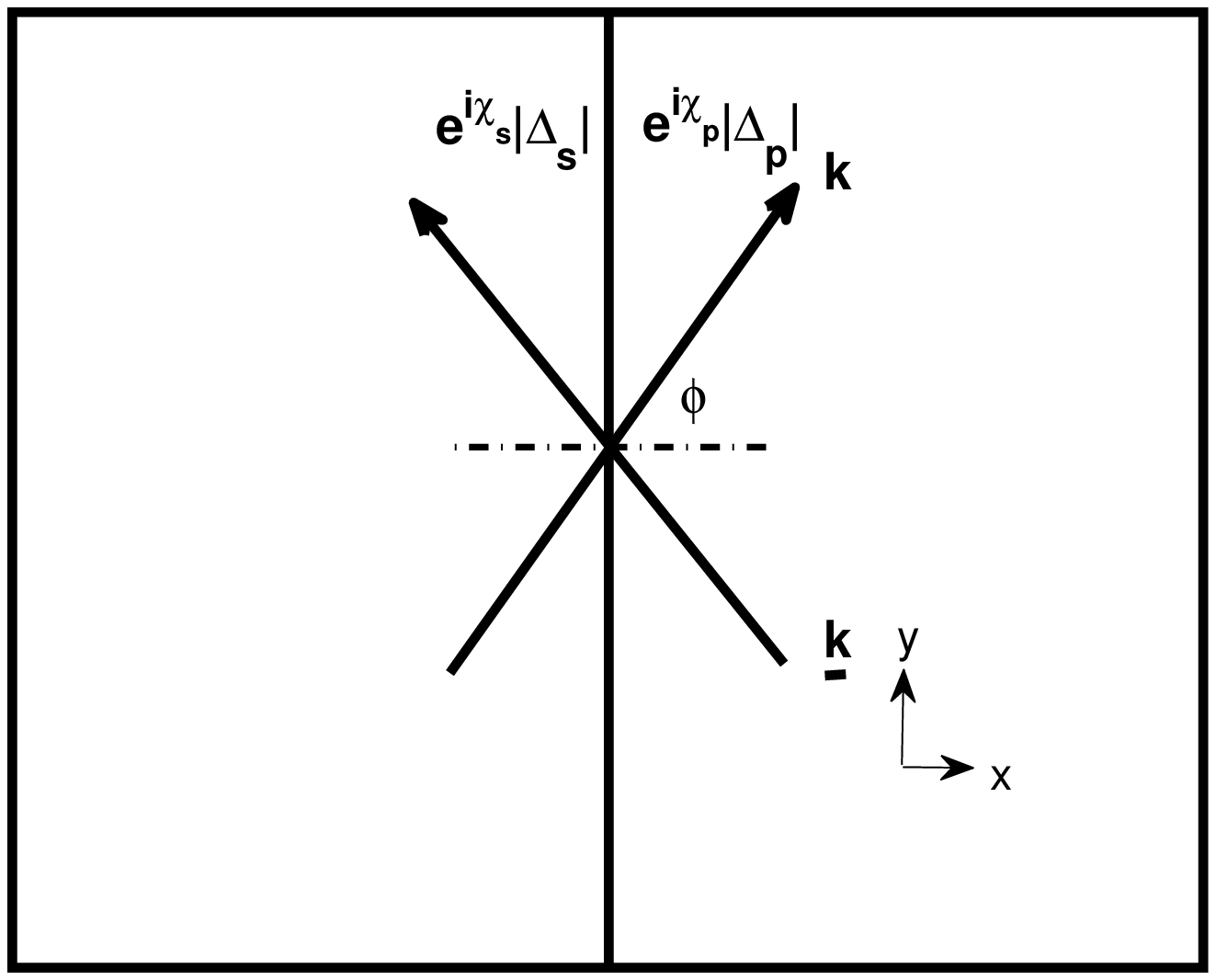}
\caption{A schematic view of the singlet-triplet junction. The
triplet superconductor, with an order parameter of magnitude
$|\Delta_p|$ and phase $\chi_p$, occupies the right (x$>$0) while
the singlet one, whose respective values denoted by $|\Delta_s|$ and
$\chi_s$, occupies the left (x$<$0). The quasiclassical path is
denoted by the direction of quasiparticle momentum $\hat{k}$. The
angle $\phi$ is defined with respect to the x-axis. Incoming and
outgoing paths labelled by $\hat{\underline{k}}$ and $\hat{k}$,
respectively, are used for interface with non-perfect
transmission.}\label{layout}
\end{figure}

We shall then consider a Josephson junction between an $s$-wave
superconductor and a purely triplet superconductor with
$\hat{d}=k_x\hat{y}-k_y\hat{x}$. For simplicity we shall consider
the two dimensional case, or equivalently the three-dimensional
case but dispersionless in $k_z$. A schematic view of the junction
is shown in Fig.\ \ref{layout}. We shall show that symmetry
argument forbids existence of certain quantities, and in the case
where a quantity is allowed, its dependence on the phase
difference is constrained. We search for symmetry operations under
which the junction would map back to itself. Caution has to be
made to account for possible changes of the phase of the order
parameters under these operations. These considerations are along
the same line as those applied earlier by one of
us\cite{Yip90,Yip93} to the Josephson current across a junction.

The s-wave (triplet) superconductor occupies $x <(>)0$. The order
parameter $\underline{\Delta}$ is a $2 \times 2$ matrix in spin
space. We have, for $x < 0$, $\underline{\Delta} = \Delta_s ( i
\sigma_y ) $ whereas for $x > 0$,  $ \underline{\Delta}(\hat k) =
\Delta_p i \left(\vec d(\hat k) \cdot \vec \sigma \right) \sigma_y$,
$\vec d(\hat k) = \hat k_x \hat y - \hat k_y \hat x$ specifies the
triplet structure of the pairs. We shall for simplicity ignore
anisotropy of the magnitude of the superconducting gaps.  In this
case, $\Delta_s$ and $\Delta_p$ are independent of $\hat k$.

First we consider the time-reversal transformation $\Theta$ under
which the supercurrent and spin accumulation are odd while the spin
current is even. The annihilation operators transform as $\Theta
a_{\vec k, \uparrow} \Theta^{-1} = a_{- \vec k, \downarrow}$ and
$\Theta a_{\vec k, \downarrow} \Theta^{-1} = - a_{- \vec k,
\uparrow}$. Using the fact that ${\underline \Delta}$ transforms as
the corresponding anomalous average, simple algebra then shows that
$\Delta_s \to \Delta_s^*$, $\Delta_p \to \Delta_p^*$ with $\hat d$
unchanged (using that $\hat d$ is real).  Hence the phase difference
changes sign.  It follows that the supercurrent $J_j (\chi) = - J_j
(-\chi)$, spin accumulation $S^i (\chi) = - S^i (-\chi)$, and spin
current $J^i_j (\chi) = J^i_j(-\chi)$ for polarization and flow
along $i$ and $j$, respectively.

Under a reflection in the x-z plane, the order parameter
$\hat{d}=k_x\hat{y}-k_y\hat{x}$ transforms according to
$(k_x,k_y,k_z)\to (k_x,-k_y,k_z)$ and
$(\hat{x},\hat{y},\hat{z})\to (-\hat{x},\hat{y},-\hat{z})$,
respectively. Hence both superconductors are invariant and the
phase difference $\chi$ is also unchanged. The only non-vanishing
currents, spins and spin-currents allowed are thus $J_{x,z}$,
$S^y$, $J^y_{x,z}$, and $J_y^{x,z}$. Since the dispersion in z is
not considered, $J_z$ and $J^y_z$ will not be mentioned hereafter.
We can also consider a reflection in the x-y plane under which
$(k_x,k_y,k_z)\to (k_x,k_y,-k_z)$ and
$(\hat{x},\hat{y},\hat{z})\to (-\hat{x},-\hat{y},\hat{z})$. The
resulting order parameter $\Delta_s \to \Delta_s$ but $\Delta_p
\to \Delta_p e^{ i \pi}$, hence the phase difference $\chi \to
\chi+ \pi$. We then have

\bea
    J_{x} (\chi) &=& J_{x} (\chi + \pi)\:, \nonumber \\\label{jsj}
    %J_{z} (\chi) &=& -J_{z} (\chi + \pi)\:, \nonumber \\
    S^y (\chi) &=& - S^y (\chi + \pi)\:, \nonumber \\
    J^y_x (\chi) &=& - J^y_x (\chi + \pi)\:, \\
    %J^y_z (\chi) &=&  J^y_z (\chi + \pi)\:,  \\
    J^x_y (\chi) &=& - J^x_y (\chi + \pi)\:, \nonumber \\
    J^z_y (\chi) &=&  J^z_y (\chi + \pi)\:. \nonumber
\eea Other symmetry operations (such as $\pi$ rotation about $\hat
x$) just produce relations that can be
found by combinations of  those listed above. %In our two-dimensional
%case, the absence of $z$-dispersion forbids $J_z$, $J_z^y$.  The
%only non-vanishing quantities allowed by symmetries are thus $J_x$,
%$S^y$, $J^y_x$, $J^x_y$ and $J^z_y$.
We note in particular that the spin accumulation lies entirely
along the $y$ direction.  In the limit of zero transmission, all
quantities are independent of $\chi$.  In this case, all spin
accumulations must vanish and the only finite spin current is
$J^z_y$. These results hold even when more general components of
the $A_{2u}$ order parameter (e.g. $k_xk_y(k_x\hat{x}-k_y\hat{y})$
in Ref.\cite{group}) are included. As we shall see later, only
$J_x$, $S^y$ and $J^z_y$ are found to be finite in our
calculations.

\section{Quasiclassical Green's function}\label{main}

We now present our calculations and the quasiclassical method.  At
positions other than the interface, the quasiclassical Green's
function $\hat{g}$, a function of momentum direction $\hat{k}$,
Matsubara frequency $\epsilon_n$ and position $\vec{r}$, obeys

\be
    [ i \epsilon_n \tau_3 - \hat \Delta, \hat g ] + i \vec v_f(\hat
    k) \cdot \vec \nabla \hat g = 0\:, \label{qc1}
\ee with the normalization condition

\be
    \hat g^2 = - \pi^2\:. \label{qc2}
\ee

Here $\vec v_f (\hat k)$ is the Fermi velocity.  The boundary
condition at $x = 0$ will be stated below.  $\hat \Delta (\hat k)$
specifies the off-diagonal pairing field.  $\hat \Delta = \left(
\begin{array}{cc} 0 & \underline{\Delta} \\ - \underline{\Delta}^\dag
& 0 \end{array} \right)$ where $\underline{\Delta}$ isthe $2 \times
2$ order parameter matrix in spin space.   With $\tau_{+} \equiv \left( \ba{cc} 0 & 1 \\
0 & 0 \ea \right)$ and $\tau_{-} \equiv \left( \ba{cc} 0 & 0 \\
1 & 0 \ea \right)$ in particle-hole space, we then have $\hat \Delta
= \Delta_s ( i \sigma_y) \tau_{+} + \Delta_s^* ( i \sigma_y)
\tau_{-}$ for $x < 0$ and $\hat \Delta = \Delta_p i ( \vec d (\hat
k) \cdot \vec \sigma) \sigma_y \tau_{+} + \Delta_p^* i \sigma_y (
\vec d (\hat k) \cdot \vec \sigma) \tau_{-}$ for $x > 0$. In order
to have tractable analytic solutions for $\hat g$, we shall also
ignore the self-consistent gap equation and hence the spatial
dependence of $\Delta_s$ and $\Delta_p$.  We shall also assume for
simplicity that the Fermi velocities magnitudes of the two
superconductors are identical and independent of $\hat k$.

\subsection{Perfect Transmission}

In this case the boundary condition at $x=0$ is simply that $\hat
g$ is continuous.  $\hat g (\hat k, \epsilon_n, 0)$ is given by
(see Appendix \ref{trick}),

\be
    \hat g (\hat k, \epsilon_n, 0) = - i \pi \{ \hat a, \hat b
    \}^{-1} [ \hat a, \hat b]\:, \label{g0}
\ee where $[,]$ and $\{,\}$ denote commutators and anti-commutators
and $\hat{a}$ and $\hat{b}$ the appropriate exponentially decaying
and increasing solutions along the quasiclassical path.

For $k_x > 0$ we need $\hat a = \hat a_p$, $\hat b = \hat b_s$ in
eq.\ (\ref{g0}). We find

\be
    \hat g(\hat k, \epsilon_n, 0) = c_3 (\hat k,\epsilon_n) \tau_3
    + c'_3 (\hat k, \epsilon_n)
    \left( \ba{cc} \hat{d}(\hat{k})\cdot\vec{\sigma} & 0 \\
    0 & -\sigma_y\hat{d}(\hat{k})\cdot\vec{\sigma}\sigma_y\ea \right)
    + ( o.d.)\:,\label{g0d1}
\ee where $(o.d.)$ denotes off-diagonal terms in particle-hole
space that we would not need,

\be
    c_3 (\hat k,\epsilon_n) = - \pi
    \frac{|\Delta_p|^2 |\Delta_s|^2 {\rm sin} \chi {\rm cos} \chi +
    i  \epsilon_n (\alpha_s \alpha_p + \epsilon_n^2)  (\alpha_p +
    \alpha_s)}
    { (\alpha_s \alpha_p + \epsilon_n^2)^2 - |\Delta_p|^2 |\Delta_s|^2 {\rm cos}^2 \chi
    }\:,\label{c3}
\ee

\be
    c'_3 (\hat k,\epsilon_n)=\pi |\Delta_p| |\Delta_s|
    \frac{(\alpha_s \alpha_p + \epsilon_n^2) {\rm sin} \chi +
    i  \epsilon_n  (\alpha_p + \alpha_s) {\rm cos} \chi}
    { (\alpha_s \alpha_p + \epsilon_n^2)^2 -  |\Delta_p|^2 |\Delta_s|^2 {\rm cos}^2 \chi }\:,\label{c3'}
\ee where $\chi \equiv \chi_p - \chi_s$ is the phase difference.
$\alpha_s \equiv \left( \epsilon_n^2 + |\Delta_s|^2
\right)^{1/2}$, $\alpha_p \equiv \left( \epsilon_n^2 +
|\Delta_p|^2 \right)^{1/2}$. The result for $c_3$ was also given
in Ref.\cite{Yip93}. For $k_x < 0$, we need $\hat{a} = \hat{b}_p$,
$\hat{b} = \hat{a}_s$ in eq.\ (\ref{g0}). Alternatively, we can
also use the symmetry \cite{Serene} $\hat g (- \hat k,
-\epsilon_n) = \tau_2 \hat g^{tr} (\hat k, \epsilon_n) \tau_2$
where $tr$ denotes the transpose. $\hat g(\hat k, \epsilon_n)$ is
still of the form in eq.\ (\ref{g0d1}), with $c_3 (-\hat k,
-\epsilon_n) = - c_3 (\hat k, \epsilon_n)$ and $c'_3 (-\hat k,
-\epsilon_n) = - c'_3 (\hat k, \epsilon_n)$. Note that we have
defined the $c_3$, $c'_3$ coefficients with $\hat k$ dependent
$\hat d$ vector in eq.\ (\ref{g0d1}), and $\hat d (-\hat k) = -
\hat d (\hat k)$.

The number current density along $x$ can in general be expressed
as

\be
  J_x = \frac{1}{2}N_f v_f \int \frac{d \phi}{2 \pi} ({\rm cos} \phi)
   T \sum_n{\rm{Tr}}
   \left[
   \tau_3\hat{g}(\hat{k},\epsilon_n)
   \right]\:,\label{current}
\ee where $\phi$ is the angle of $\hat k$ with respect to $\hat
x$, $N_f$ is the density of states per unit area for a single spin
species. The symbol $\rm{Tr}$ represents taking a full trace in
both the spin and particle-hole spaces. Only the $c_3$ component
in eq.\ (\ref{c3}) contributes to $J_x$. The spin density in the
$i$ direction at $x=0$ can be expressed as\cite{footnote}

\be
  S^i = \frac{\hbar}{4}N_f\int\frac{d\phi}{2\pi}
   T\sum_n{\rm{Tr}}
   \left[
   \hat{\sigma}^i\hat{g}(\hat k, \epsilon_n)
   \right]\:.\label{magnetization}
\ee Here we define the symbols $\hat{\sigma}^i$ by $\hat \sigma^x
\equiv \sigma_x$, $\hat{\sigma}^y=\sigma_y\tau_3$ and $\hat
\sigma^z \equiv \sigma_z$.  So here only $S^y$ is finite and is
associated with $c'_3$ in eq.\ (\ref{c3'}). The spin current
densities, with superscript(subscript) denoting the spin (flow)
direction at $x= 0$ is\cite{footnote}

\be
  J^i_j=\frac{\hbar}{4}N_f v_f\int\frac{d\phi}{2\pi} (\hat k_j)
  T\sum_n{\rm{Tr}}
  \left[
  \tau_3\hat{\sigma}^i\hat{g}(\hat k, \epsilon_n)
  \right]\:.
  \label{spincurrent}
\ee Note that the three components of $\hat{\sigma}^i \tau_3$ are
$ \sigma_x \tau_3$, $\sigma_y$ and $\sigma_z \tau_3$. It follows
that all the spin currents vanish since $\hat{g}$ of eq.\
(\ref{g0d1}) does not contain any $\hat{\sigma}^i\tau_3$
components. Physically, the Andreev equation for each $\hat{k}$ is
decoupled from other paths, and hence can be block-diagonalized
using quantization axis along $\hat{d}(\hat{k})$. Along this axis,
both the singlet and triplet superconductors consist of only
$\uparrow\downarrow$ pairs. These Cooper pairs do not have any net
spins, and they cannot contribute to any dissipationless spin
current.  See also the discussions near the end of subsection C.

\begin{figure}
\input{epsf}
\epsfxsize=6in \epsfysize=3in \epsfbox{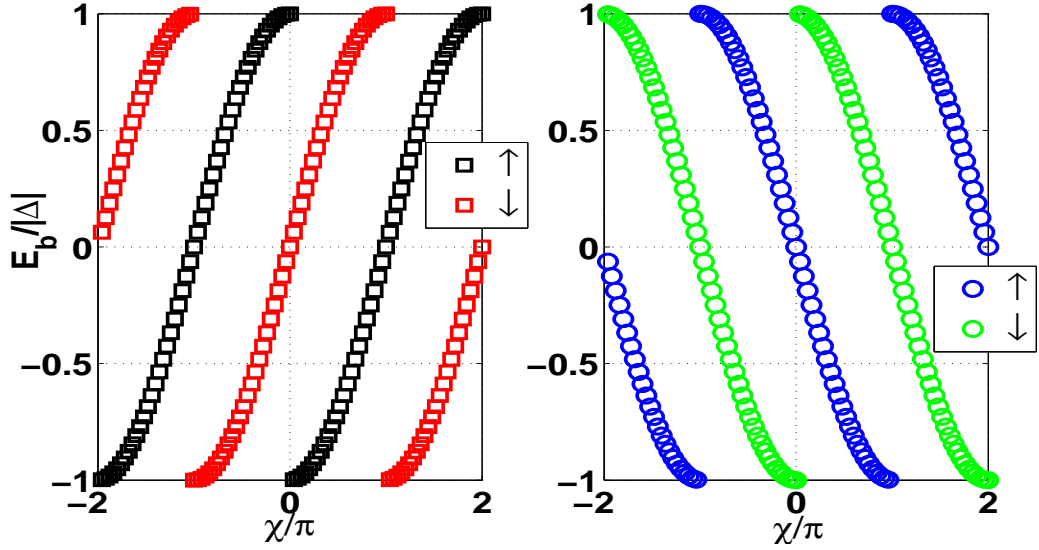}
\caption{(Color online)\ The interface bound states associated with
the $k_x>0$ and $k_x<0$ paths in the perfect transmission case. To
facilitate the discussions for spin accumulation and spin current,
$\uparrow$ spin means parallel to $\hat{d}(\hat{k})$ associated with
the right moving path, i.e. $k_x>0$, but antiparallel to
$\hat{d}(\hat{k})$ if $k_x<0$.}\label{Eb}
\end{figure}

Next we present explicit results for the case of equal gaps on
both sides, i.e. $|\Delta_s|=|\Delta_p|=|\Delta|$. Here the
interface bound states, which correspond to the poles of $\hat{g}$
in eq.\ (\ref{g0d1}), are essential for the quantities in eq.\
(\ref{current}) and (\ref{magnetization}). It can be shown that
for the right moving path ($k_x>0$), the  bound states of spin
parallel and antiparallel with $\hat{d}(\hat{k})$ are given by
$E_{b,\uparrow}=-|\Delta|\cos(\frac{\chi}{2})\rm{sgn}\left[\sin(\frac{\chi}{2})\right]$
and
$E_{b,\downarrow}=|\Delta|\sin(\frac{\chi}{2})\rm{sgn}\left[\cos(\frac{\chi}{2})\right]$,
respectively. For the left moving path ($k_x<0$), the bound state
energies are
$E_{b,\uparrow}=-|\Delta|\sin(\frac{\chi}{2})\rm{sgn}\left[\cos(\frac{\chi}{2})\right]$
and
$E_{b,\downarrow}=|\Delta|\cos(\frac{\chi}{2})\rm{sgn}\left[\sin(\frac{\chi}{2})\right]$.
Notice that we adopt a common spin quantization axis for both
right and left moving paths (caption of Fig.\ \ref{Eb}) to
facilitate the following discussions. The bound state spectra are
plotted as a function of phase difference $\chi$ in Fig.\
\ref{Eb}. It can be seen that, for a given path, the two branches
of opposite spin projections are identical except separated by
$\pi$, which reflects the invariance of triplet order parameter
under $\chi_p\rightarrow\chi_p+\pi$ and
$\hat{d}\rightarrow{-\hat{d}}$.

\begin{figure}
\input{epsf}
\epsfxsize=6in \epsfysize=6in \epsfbox{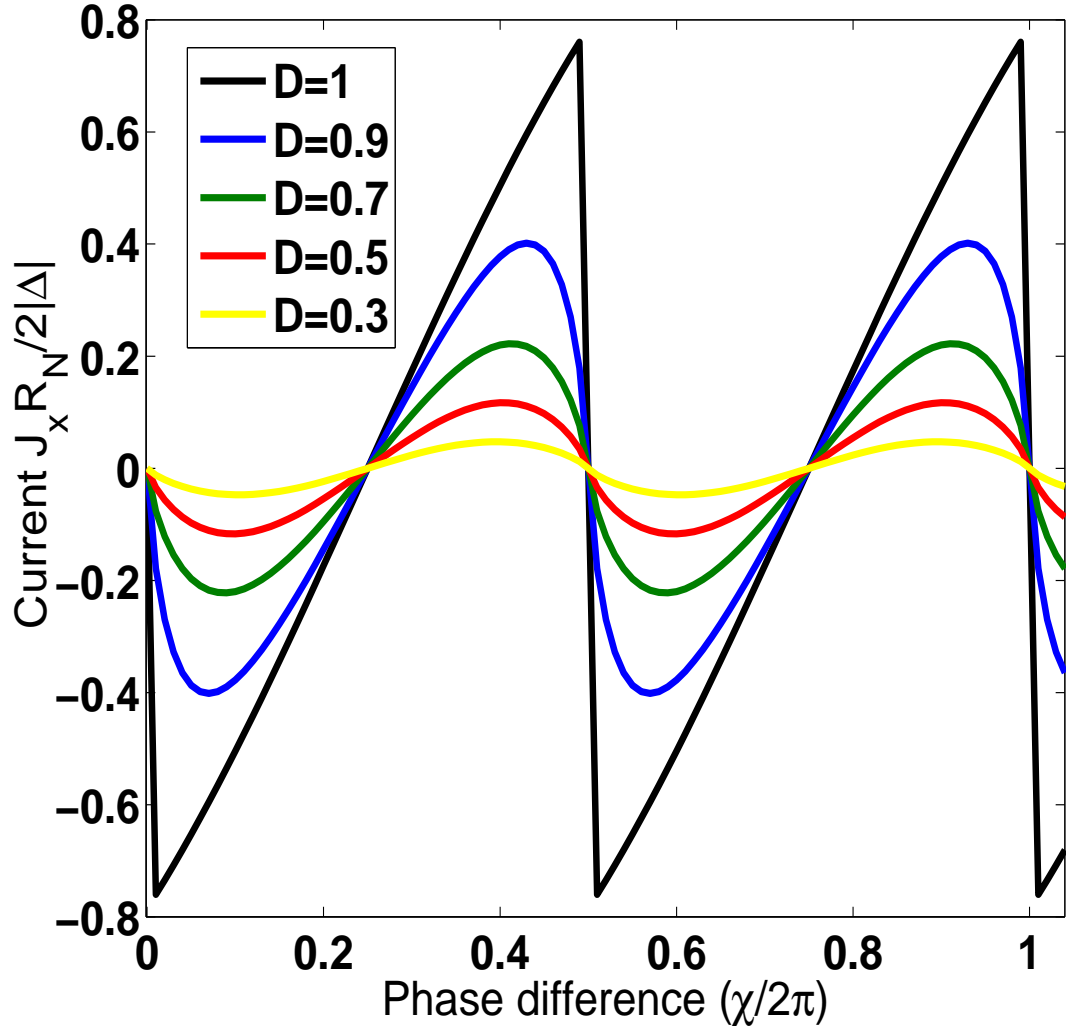}
\caption{(Color online)\ The calculated supercurrent $J_x$ for
various transmission coefficients $\mathcal{D}$. The gap on both
sides are set to equal to $|\Delta|$ and
$T=|\Delta|/100$.}\label{Jx}
\end{figure}

The analytical results for $J_x$ is obtained using eq.\
(\ref{current}), which gives

\be
    J_x=-\frac{2|\Delta|}{e^2R_N}\frac{\pi}{4}
    \left[\cos(\chi/2)\tanh\frac{|\Delta|\sin(\chi/2)}{2T}
    -\sin(\chi/2)\tanh\frac{|\Delta|\cos(\chi/2)}{2T}\right]\:,\label{jjx}
\ee where $R_N$ denotes the corresponding resistance in the normal
state. Eq.\ (\ref{jjx}) coincides with the previous results in
Ref.\cite{Yip93}. $J_x$ is plotted in Fig.\ \ref{Jx} and the
present case corresponds to the line denoted by $\mathcal{D}=1$.
$J_x$ can be understood by summing over contributions
$\frac{\partial{E_b}}{\hbar\partial\chi}$ from occupied bound
states. Notice that a current jump occurs whenever $\chi$ is a
multiple of $\pi$. When $\chi$ is slightly larger than 0, the
state labelled the red square in the left panel and the one
labelled by the green circle in the right are occupied. Only the
latter bound state with a negative slope contributes to $J_x$.
When $\chi$ is slightly less than 0, on the other hand, the black
square in the left panel with a positive slope is occupied and
contributes to $J_x$.

Moreover, the splitting between bound states actually contributes
to a finite spin accumulation near the interface along some
direction. Consider $0<\chi<\pi$ and zero temperature. Referring
to Fig.\ \ref{Eb}, for the right and left moving paths, the states
with spin parallel to the quantization axis defined in the caption
are both populated. As the parameter $\phi$ varies between
$\pm\pi/2$, this quantization axis varies. A net spin is generated
along the positive y-axis, whereas the x component adds to zero.
Analytically, the spin accumulation can be obtained from eq.\
(\ref{magnetization}), which gives

\be
    S^y=\hbar{N_f}|\Delta|
    \left[\cos(\chi/2)\tanh\frac{|\Delta|\sin(\chi/2)}{2T}
    +\sin(\chi/2)\tanh\frac{|\Delta|\cos(\chi/2)}{2T}\right]\:.
\ee As a function of $\chi$, the spin accumulation $S^y$ for both
sides of the interface is plotted in Fig.\ \ref{Sy}. The present
case corresponds to the line with $\mathcal{D}=1$. In addition,
$S^y$ is also continuous across the interface for perfect
transmission. We note however that, if the magnitude of the gaps of
the two superconductors are unequal, there can also be contributions
due to continuum states, as in the case of supercurrent between two
unequal gap s-wave superconductors.\cite{unequal} Since the Green's
function decays as $e^{ - 2 \alpha |x|/v_f | {\rm cos} \phi |}$,
$S^y$
%depends on the position as the function
%$\int d \phi ({\rm cos} \phi) e^{ - 2 \alpha x / v_f |{\rm cos} \phi|} $,
decays in a distance of order of coherence length
$\hbar{v_f}/|\Delta|$ away from the interface. The total spin
accumulation is of order $\hbar^2 N_f v_f$ per unit length along the
junction.

\subsection{No transmission}\label{noD}

In this case all particles are reflected.  The behavior of the
s-wave superconductor for $x < 0$ is trivial  and we shall thus
concentrate only on the triplet superconductor on the right. Let
us denote the incoming wavevectors by $\underline{\hat k}$ and the
reflected outgoing wavevectors by $\hat k$, with  $\hat k_x > 0$
and $\underline{\hat k}_x < 0$. See Fig.\ \ref{layout}.  We label
positions along the quasiparticle path consisting of each pairs of
$\hat k$ and $\underline{\hat k}$ by $u$, with $u< 0$ ($u > 0$)
labels the part for $\hat k$ ($\underline{\hat k}$). $\hat g(u)$
is continuous at $u=0$, and can be obtained from eq.\ (\ref{g0})
with $\hat a \to \hat a_p(\hat k)$ and $\hat b \to \hat
b_p(\underline{\hat k})$. Since $\hat d (\hat k) \neq \hat
d(\underline{\hat k})$, we shall introduce the quantities $C
\equiv \hat d (\hat k) \cdot  \hat d (\underline{\hat k})$ and
$\vec D  \equiv \hat d (\hat k) \times \hat d (\underline{\hat
k})$.  Note that $C^2 + |\vec D|^2 = 1$. The part of $\hat g(0)$
which is diagonal in particle-hole space and even in $\epsilon_n$
is found to be

\bdm
    \pi \frac{|\Delta_p|^2}{\left[ 2 \epsilon_n^2 + |\Delta_p|^2
    \left( 1 + C \right) \right]}
    \left( \ba{cc} (\vec D  \cdot \vec \sigma) & 0 \\
    0 & \sigma_y (\vec D  \cdot \vec \sigma) \sigma_y \ea \right)\:.
\edm For our state, $C = - {\rm cos} 2 \phi$ and $\vec D = \hat z
{\rm sin} 2 \phi $. It follows that there are no currents $J_j$. For
a given pair of wavevectors $\hat{k}$ and $\underline{\hat{k}}$,
there is in general a finite spin along $\vec{D}\parallel\hat{z}$.
However, the contribution from the pairs of wavevectors in opposite
directions sum to zero (That is, between the pair with outgoing
$\hat k$ and incoming wavevector being $-\hat k$, or alternatively,
$\pm \phi$). Therefore $S^z = 0$, and there is no spin accumulation
in any direction, which can also be seen by noting that $\hat{g}$
does not contain any $\hat{\sigma}^i$ component. The only finite
spin current is $J^z_y$ associated with the $ \sigma_z \tau_3$
component, and its value at $x=0$ is given by

\be
    J^z_y  = \hbar N_f v_f \int_{-\frac{\pi}{2} < \phi < \frac{\pi}{2}}
    \frac{d \phi}{\pi} \left( {\rm sin} \phi \right) T \sum_n
    \pi \frac{|\Delta_p|^2 D^z}{\left[ 2 \epsilon_n^2 + |\Delta_p|^2 \left( 1 + C \right)
    \right]}\:,
    \label{SzyD0}
\ee where $D^z$ is the $z$ component of $\vec D$.  Since $\hat
g(\hat k, \epsilon_n, 0) = \hat g(\underline{\hat k}, \epsilon_n,
0)$, the angular integral in eq (\ref{SzyD0}) has been replaced by
twice the contribution due to outgoing wavevectors. The factor ${\rm
sin} \phi$ is due to $\hat k_y = \underline{\hat k}_y$. At zero
temperature, the spin current density
$J^z_y={\hbar}N_fv_f\frac{|\Delta_p|}{2}$ at the interface and
decays into the bulk within a coherence length.  The total spin
current is of order $\hbar^2 N_f v_f^2$.

The physical picture of the spin-current is similar to that of the
edge current in the so-called chiral
superconductors,\cite{helicalSC} and has been discussed also in,
e.g., Ref.\cite{Vorontsov}. Our triplet state consists of
$\uparrow\uparrow$ pairs and $\downarrow\downarrow$ pairs only,
with wavefunctions respectively given by
$(-d^x+id^y)\to{i}(k_x-ik_y)$ and $(d^x+id^y)\to{i}(k_x+ik_y)$.
Due to the phase difference between the order parameters of the
incoming and outgoing momenta, each spin component has a bound
state (for a given pair of incident and reflected
wave-wavevectors) but opposite energies: $\epsilon=\mp
|\Delta_p|\sin\phi$ for spin up (down) respectively near the
surface. Thus the up (down) spins preferentially occupy the states
with positive (negative) $y$ momentum, contributing to a net spin
current $J^z_y$.  In this picture, it also follows that $J^x_y$
vanishes for $\mathcal{D} = 0$.

\subsection{General Transmission}

In this subsection we consider a general interface between our
singlet and triplet superconductor of (angular and spin independent)
transmission coefficient $\mathcal{D}$. We denote the incoming
(outgoing) wavevector on the right by $\hat {\underline k}$ and
$\hat k$, and conversely for the left, see Fig.\ \ref{layout}.  The
corresponding Green's functions
$\hat{g}(\hat{\underline{k}},x=0_{\pm})$,
$\hat{g}(\hat{k},x=0_{\pm})$ on the two sides of the spin-inactive
interface are related to each other by a set of boundary conditions
given in Ref.\cite{Yip97}. See also Appendix \ref{trick}. It is more
convenient to express them in terms of the difference
$\hat{g}_d=\hat{g}(\hat{k},x=0_{+})-\hat{g}(\hat{\underline{k}},x=0_{+})
=\hat{g}(\hat{k},x=0_{-})-\hat{g}(\hat{\underline{k}},x=0_{-})$,
which is continuous across the interface, and the sums
$\hat{s}^{r(l)}=\hat{g}(\hat{k},x=0_{+(-)})+\hat{g}(\hat{\underline{k}},x=0_{+(-)})$.
It can be shown that the supercurrent $J_x$ in eq.\ (\ref{current})
and spin currents $J^i_x$ in eq.\ (\ref{spincurrent}) across the
interface can be expressed solely in terms of the difference
$\hat{g}_d$. Note that $\phi$ which specifies the angle for $\hat k$
is now restricted within $\pm\pi/2$. Moreover, the $\tau_3$ and
$\tau_3\hat{\sigma}^i$ components of $\hat{g}_d$ are associated with
$J_x$ and $J^i_x$, respectively. Below we consider the equal gap
case for simplicity in which $\hat{g}_d$ can be worked out
analytically via eq.\ (\ref{gd}). The $\tau_3$ component
contributing to $J_x$ is given by

\be
    \left[\hat{g}_d(\hat{k},\epsilon_n)\right]_{\tau_3}=
    \frac{
    (-\pi)%|\Delta|^2
    \mathcal{D}^2|\Delta|^4\sin(2\chi)%\ \tau_3
    %+4i\mathcal{D}\alpha\epsilon_n\cos\phi\cos\chi
    %\ \sigma_2\tau_3
    }
    {
    4\alpha^2\epsilon_n^2
    +\mathcal{D}^2|\Delta|^4\sin^2\chi
    +4(1-\mathcal{D})\alpha^2|\Delta|^2\sin^2\phi
    }\:,\label{gdD}
\ee By numerically performing the sum over the Matsubara
frequencies, $J_x$ for arbitrary $\mathcal{D}$ is plotted in Fig.\
\ref{Jx}. Note that the current is odd and is periodic in the
phase difference $\chi$ with period $\pi$, as noted also in Ref.\
\cite{Yip93}. See also Sec.\ref{symmetry}. Second, we find that
none of the $\tau_3\hat{\sigma}^i$ components appear in
$\hat{g}_d$, and hence all the spin currents $J^i_x$ across the
junction are zero. We note, however, that the (spatial) symmetry
argument in Sec.\ \ref{symmetry} allows a nonzero $J^y_x$ as in
eq.\ (\ref{jsj}). Therefore, the vanishing of $J^y_x$ results from
other symmetries which we shall discuss near the end of this
section.

\begin{figure}
\input{epsf}
\epsfxsize=6in \epsfysize=3in \epsfbox{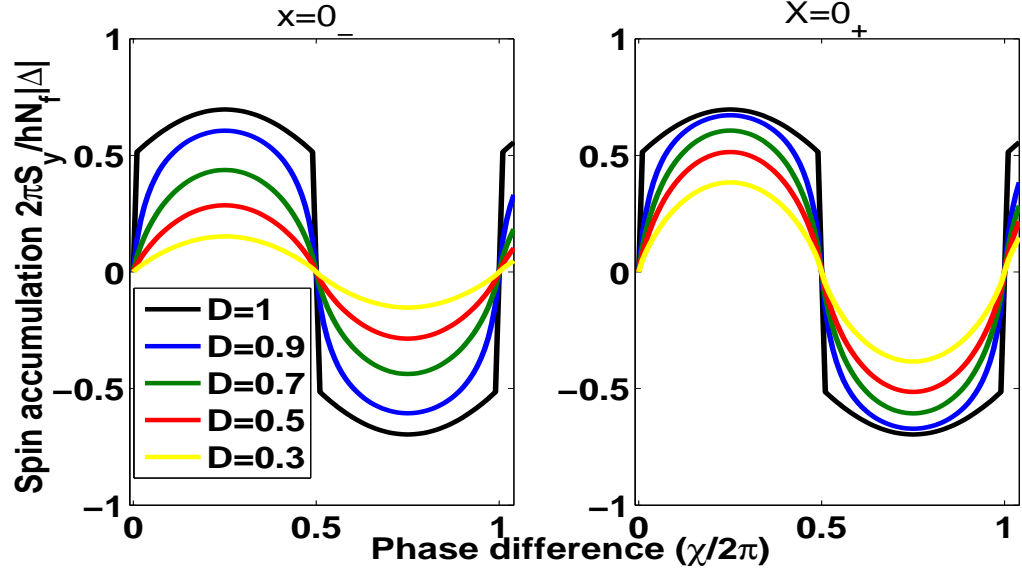}
\caption{(Color online)\ The spin accumulation $S^y$ at $x =
0_{\mp}$ on the two sides of the interface for various transmission
coefficients $\mathcal{D}$ with $T=|\Delta|/100$.}\label{Sy}
\end{figure}

\begin{figure}
\input{epsf}
\epsfxsize=6in \epsfysize=6in \epsfbox{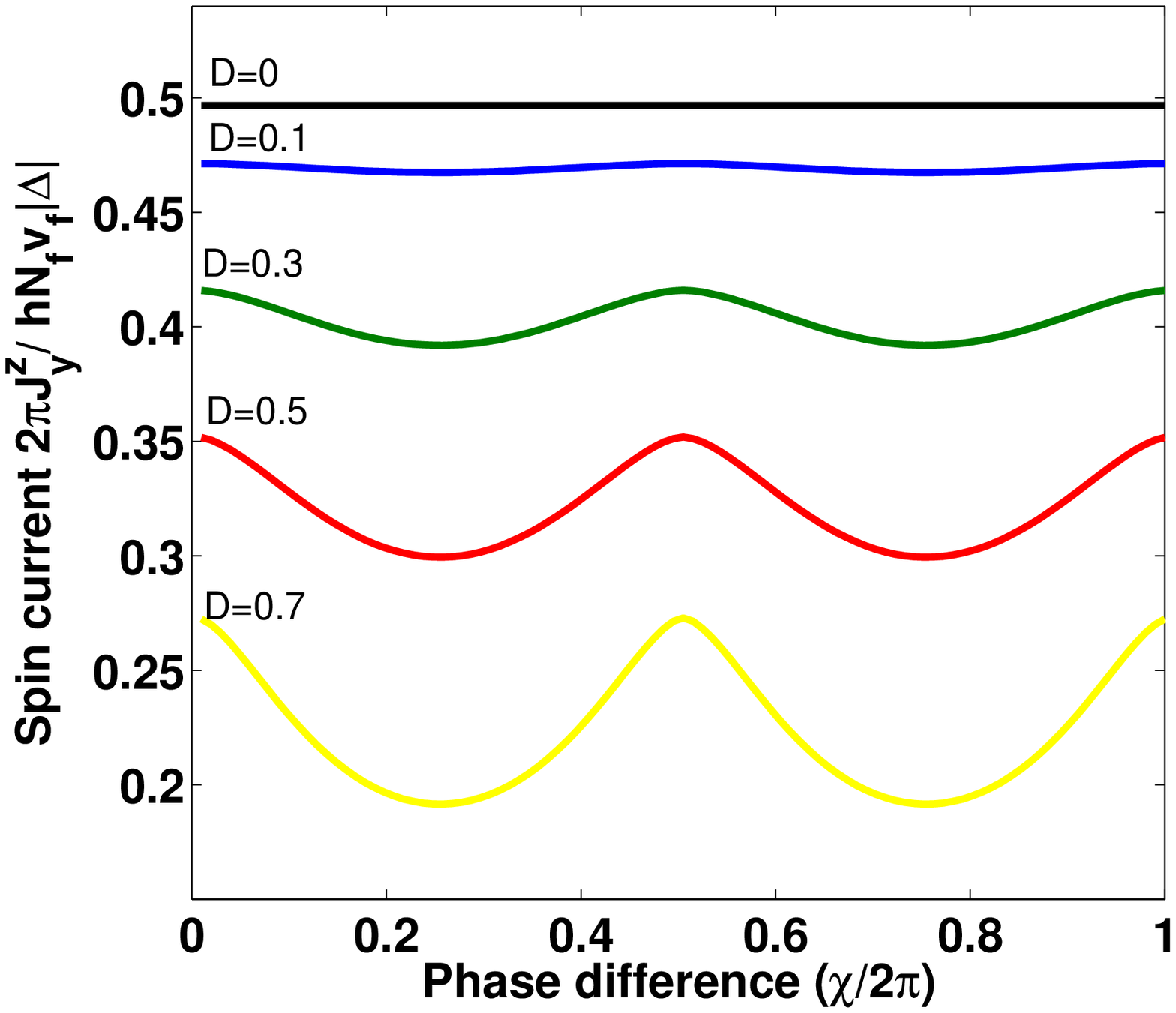}
\caption{(Color online)\ The spin current $J^z_y$ at $x = 0_{+}$.
 T=$|\Delta|/100$ here. On the
singlet side of interface, the spin current is zero for all
$\mathcal{D}$.}\label{Jzy}
\end{figure}

At the right side of the interface, the spin accumulation $S^i$ and
the spin current $J^i_y$ flowing parallel to the interface can all
be expressed in terms of $\hat{s}^r$ solely. Here the
$\hat{\sigma}^i$ components in $\hat{s}^r$ are needed for $S^i$ and
the $\tau_3\hat{\sigma}^i$ ones are for $J^i_y$ as required in eq.\
(\ref{magnetization}) and (\ref{spincurrent}). By using eq.\
(\ref{sr}), the $\hat{\sigma}^i$ components are listed below,

\be
    \left[\hat{s}^r(\hat{k},\epsilon_n)\right]_{\hat{\sigma}^i}=
    4\pi\mathcal{D}|\Delta|^2
    \frac{-i \sin\phi\cos\chi\alpha\epsilon_n\ \sigma_x
    + \left[(1-\frac{\mathcal{D}}{2})\alpha^2+\frac{\mathcal{D}}{2}\epsilon_n^2\right]
    \cos\phi\sin\chi\ \sigma_y\tau_3}
    {4\alpha^2\epsilon_n^2
    +\mathcal{D}^2|\Delta|^4\sin^2\chi
    +4(1-\mathcal{D})\alpha^2|\Delta|^2\sin^2\phi}\:.
    \label{srS}
\ee The spin accumulation $S^x$ is identically zero because the
coefficient in $\sigma_x$ is odd in $\epsilon_n$, and the factor
factor $\sin\phi$ also gives zero after the angular integration.
This result is consistent with our symmetry argument in Sec
\ref{symmetry}. The only finite spin accumulation is $S^y$ which is
shown in Fig.\ \ref{Sy} due to the $\sigma_y \tau_3$ component in
eq.\ (\ref{srS}). Note that $S^y(\chi)$ obeys the symmetry in Sec
\ref{symmetry} and has period $ 2 \pi$. As for the spin current, the
only nonvanishing component of $\tau_3\hat{\sigma}^i$ is given by,

\be
    \left[\hat{s}^r(\hat{k},\epsilon_n)\right]_{\tau_3\hat{\sigma}^i}=
    \frac{4\pi|\Delta|^2(1-\mathcal{D})\alpha^2\sin(2\phi)\ \sigma_z\tau_3}
    {4\alpha^2\epsilon_n^2
    +\mathcal{D}^2|\Delta|^4\sin^2\chi
    +4(1-\mathcal{D})\alpha^2|\Delta|^2\sin^2\phi}
    \:.\label{SrSC}
\ee For $\mathcal{D}=0$, $J^z_y$ does not depend on $\chi$. For
$\mathcal{D}<1$, the phase dependence comes from the $\sin^2\chi$
term in the denominator. The $J^z_y$ versus the phase difference
$\chi$ is plotted in Fig.\ \ref{Jzy} for various $\mathcal{D}$. This
spin current is even in $\chi$ and is periodic with period $\pi$.
(see Sec \ref{symmetry}). We note that the vanishing of
$\tau_3\hat{\sigma}^x$ components leads to zero $J^x_y$ which was
not anticipated by our symmetry argument in Sec.\ \ref{symmetry}.

At the left side of interface, $S^i$ and $J^i_y$ can be calculated
via $\hat{s}^l$ in eq.\ (\ref{sl}). The $\hat{\sigma}^i$
components are identical to those in $\hat{s}^r$ except that
$\alpha$ and $\epsilon_n$ are interchanged in the square bracket
$[...]$ of eq.\ (\ref{srS}) associated with $\sigma_y\tau_3$
component. The numerical results for $S^y(x=0_-)$ are also shown
in Fig.\ \ref{Sy}. It can be seen that $S^y$ is continuous across
the interface only for $\mathcal{D}=1$. In addition, all the terms
associated with the spin current $J^i_y$, including the
$\sigma_z\tau_3$ component, vanish for all $\mathcal{D}$.
Consequently, all the spin currents vanish on the left side.

The vanishing of $J^i_j$ for $x < 0$ and $J^i_x$ for all $x$ is a
result of spin-conservation.  Observing that $\sigma_x \tau_3$,
$\sigma_y$, $\sigma_z \tau_3$ commute with $\tau_3$ and $\hat
\Delta_s$, we see that, by multiplying eq.\ (\ref{qc1}) by these
matrices and then taking the trace, $ \vec v_f \cdot \nabla \left(
{\rm Tr} [ \hat \sigma^i \tau_3 \hat g ] \right) = 0$.  That is,
the spin current is constant along any quasclassical path at any
point inside the singlet superconductor.  Since the spin-current
vanishes at $x \to - \infty$, it follows that the spin current on
each quasiclassical path vanishes for $x< 0$.  Hence $J^i_j = 0$
for all $i$, $j$ if $x < 0$.  Note that this vanishing of the
spin-current does not rely on angular integration.  Since $J^i_x =
0$ for $x = 0_{-}$ and the spin current is continuous across an
spin-inactive interface ($\hat g_d$ is continuous), $J^i_x = 0$
also for $x = 0_{+}$.  At any point $x > 0$, the Green's function
is a linear combination of its value at $x = 0_{+}$ and $x \to
\infty$ where $J^i_x $ also vanishes.  Hence $J^i_x = 0$ also for
all $x > 0$. The vanishing of $J^i_j$ at $x = 0_{-}$ and $J^i_x$
at $x=0$ can also be easily proven using eq.\ (\ref{gd}) and
(\ref{sl}) using that fact that $\hat \sigma^i \tau_3$ commutes
with $\hat g^l_{\rm aux}$ and ${\rm Tr} [\hat \sigma^i \tau_3 \hat
g^l_{\rm aux}] = 0$.

As mentioned, the symmetry allowed $J^x_y$ is found to vanish in
our calculation. We have checked that the vanishing of $J^x_y$ is
also true in the case of $|\Delta_s|\neq|\Delta_p|$.  We do yet
not have a simple physical explanation of this result.
Mathematically, this follows from the fact that absence of the
$\sigma_x \tau_3$ term for $\hat g$ on the left of the interface
(due to spin conservation) is carried over to $\hat g$ on the
right. Vorontsov \emph{et al}.\cite{Vorontsov} have also
considered the interface between vaccum and a noncentrosymmetric
superconductor with finite spin-orbital Rashba energy, which lifts
the energy degeneracy between quasiparticles at the same momentum
but opposite spin projections. They showed that this can lead to
some finite and oscillating $J^x_y$ and $J^y_x$. We expect that
this may also happen in our junction.

\section{Discussions and conclusions}\label{summary}

We have considered the spin accumulation and spin current near a
Josephson junction between a singlet and triplet superconductor. We
showed that symmetry arguments (Sec.\ \ref{symmetry}) place strong
restrictions on the existence of above physical quantities and their
dependence on phase difference $\chi$ across the Josephson junction.
Comparing with the pervious work,\cite{Sengupta} this method also
applies for any triplet pairing wavefunction and provides a more
general way of determining the direction in which the spin lies.
Conversely, the direction and phase dependence of the spin
accumulation actually inform us about which symmetry is broken by
the junction and hence the symmetry of the triplet order parameter
itself. Moreover, the quasiclassical Green's function technique is
employed to quantitatively investigate the predicted supercurrent
$J_x$, spin accumulation $S^y$, and spin current $J^{x,z}_y$.
$J^x_y$ turns out to be zero for our junction, though it is symmetry
allowed. For transmission coefficient $0<\mathcal{D}<1$ in our
calculation, the spin accumulation $S^y$ and spin current $J^z_y$
coexist within a coherence length at the triplet side, a feature
which does not appear in the previous
studies.\cite{Vorontsov,Sengupta}

In conclusions, we have calculated the spin accumulation and spin
current near the interface of a singlet-triplet junction with the
triplet order parameter specified by
$\hat{d}=k_x\hat{y}-k_y\hat{x}$. The method of quasiclassical
Green's functions as well as the symmetry arguments can be
generalized to other junction with arbitrary pairing symmetries.
These spin accumulation and dissipationless spin currents depends on
the phase difference and hence can be controlled by the charge
current passing through the junction.

\section{Acknowledgement}

This research was supported by the National Science Council of
Taiwan under grant number NSC-95-2112-M001-054-MY3.

\appendix
\section{exploding \& decaying trick}\label{trick}

In this Appendix we explain the exploding and decaying trick. This
trick has been used for pure s-wave\cite{Thuneberg} and pure
p-wave pairing (e.g.,\ Ref.\cite{Fogelstrom}).  From these
references, one can actually deduced that the method can be
generalized to mixed singlet and triplet pairs, so that results
such as eq.\ (\ref{g0}) can still be used.  However, we would like
to provide our alternate derivation below to show that it is
indeed applicable for mixed pairing, and moreover we believe that
our presentation may be more transparent to some readers than
those in the literature.  We also note that this method is not
limited to spatial independent gaps, though we shall discuss only
the (piecewise) constant gaps case to simplify our presentation.
Furthermore, this method can be easily implemented numerically, as
has been performed in, e.g., \cite{Thuneberg,Fogelstrom,Yip97}
etc.

We begin by reviewing the first the trick for pure s-wave
superconductor. Writing $u$ as the parameter along a
quasiclassical path, eq.\ (\ref{qc1}) can be written as

\be
\left[ i \epsilon_n \tau_3 - \Delta_s ( i \sigma_y) \tau_{+}-
 \Delta_s^* ( i \sigma_y) \tau_{-}, \ \hat g(u) \right]
  + i v_f \partial_u \hat g(u) \ = \ 0
  \label{qcs}
\ee where we have suppressed the $\hat k$ and $\epsilon_n$
dependence of $\hat g$.  A "constant" solution (satisfying also
eq.\ (\ref{qc2})), which is also the $\hat g$ for a bulk
superconductor, is given by

\be
    \hat g_{s,{\rm{bulk}}} = - \pi \frac{i \epsilon_n \tau_3 -
    \Delta_s ( i \sigma_y) \tau_{+}- \Delta_s^* ( i \sigma_y)
    \tau_{-}}{\left( \epsilon_n^2 + |\Delta_s|^2 \right)^{1/2}}
\ee and is thus a linear combination of $\tau_3$,$\sigma_y \tau_+$
and $\sigma_y \tau_{-}$ matrices only. It is also possible to find
other solutions to eq.\ (\ref{qc1}) (without satisfying eq.\
(\ref{qc2})) which are linear combination of these three matrices
only. They are, with $\alpha_s \equiv \left( \epsilon_n^2 +
|\Delta_s|^2 \right)^{1/2}$,

\bea
    \hat a_s(u) &=& e^{- 2 \alpha_s u / v_f}
    \left( - i |\Delta_s|^2 \tau_3 - \Delta_s ( \alpha_s + \epsilon_n) i \sigma_y \tau_{+} +
    \Delta_s^* (\alpha_s - \epsilon_n)  i \sigma_y \tau_{-} \right) \\
    \hat b_s(u) &=& e^{+ 2 \alpha_s u / v_f}
    \left( + i |\Delta_s|^2 \tau_3 - \Delta_s ( \alpha_s - \epsilon_n) i \sigma_y \tau_{+} +
    \Delta_s^* (\alpha_s + \epsilon_n)  i \sigma_y \tau_{-} \right)
\eea which will be called the decaying and exploding solutions "in
the same block"\cite{Fogelstrom}. We note that they satisfy $\hat
a^2 = $, $\hat b^2 = 0$, $\{ \hat g_{s,{\rm bulk}}, \hat a \} = \{
\hat g_{s,{\rm bulk}}, \hat b \} = 0$. In fact, $\hat g_{s,{\rm
bulk}}$ can be written  as $\hat g = - i \pi ( \hat P_1 - \hat P_2
)$ with $\hat P_1 = \hat a \hat b / \{\hat a, \hat b\}$ and $\hat
P_2 = \hat b \hat a / \{\hat a, \hat b\}$ being projection
operators with $\hat P_1 + \hat P_2 = 1$, $\hat P_1 \hat P_2 =
\hat P_2 \hat P_1 = 0$, and $\hat a \hat P_1 = 0$, $\hat a \hat
P_2 = \hat a$, $\hat b \hat P_1 = \hat b$, $\hat b \hat P_2 = 0$
(see, e.g., Ref.\cite{Yip97}).

Similar results apply to the pure triplet superconductor.  The bulk solution is

\be
    \hat g_{p,\rm{bulk}} = - \pi \frac{i \epsilon_n \tau_3 -
    \Delta_p ( i \vec d \cdot \vec \sigma \sigma_y) \tau_{+}-
    \Delta_p^* ( i \sigma_y \vec d \cdot \vec \sigma) \tau_{-}}
    {\left( \epsilon_n^2 + |\Delta_p|^2 \right)^{1/2}}
\ee and is thus a linear combination of $\tau_3$, $(\vec d \cdot
\vec \sigma) \sigma_y \tau_+$ and $\sigma_y (\vec d \cdot \vec
\sigma) \tau_{-}$ matrices only. The other solutions to eq.\
(\ref{qc1}) (without satisfying eq.\ (\ref{qc2})) which are linear
combination of these same three matrices are, with $\alpha_p
\equiv \left( \epsilon_n^2 + |\Delta_p|^2 \right)^{1/2}$:

\bea
    \hat a_p(u) &=& e^{- 2 \alpha_p u / v_f}
    \left( - i |\Delta_p|^2 \tau_3 -
    \Delta_p ( \alpha_p + \epsilon_n) i (\vec d \cdot \vec \sigma) \sigma_y \tau_{+} +
    \Delta_p^* (\alpha_p - \epsilon_n)  i \sigma_y (\vec d \cdot \vec \sigma) \tau_{-} \right) \\
    \hat b_p(u) &=& e^{ + 2 \alpha_p u / v_f}
    \left( + i |\Delta_p|^2 \tau_3 -
    \Delta_p ( \alpha_p - \epsilon_n) i (\vec d \cdot \vec \sigma) \sigma_y \tau_{+} +
    \Delta_p^* (\alpha_p + \epsilon_n)  i \sigma_y (\vec d \cdot \vec \sigma) \tau_{-} \right)
\eea

Let us now consider our junction, and begin with the case of
perfect transmission. $\hat \Delta = \Delta_{s,p}$ for $x < (>)
0$, and $\hat g$ is continuous at $x = 0$. Let us first consider
$k_x > 0$, and label a point on the quasiclassical path by $u$,
with $u= 0$ at the interface. (hence $u = x / \hat k_x$).  $\hat
g$ must decay to $\hat g_{s, \rm bulk}$ ($\hat g_{p, \rm bulk}$)
as $u \to -\infty$ ($+\infty$).  We note however, that we cannot
just try the ansatz $\hat g(u) = \hat g_{p,\rm  bulk} + c_p \hat
a_p (u)$ for $u > 0$ and $\hat g(u) = \hat g_{s,\rm  bulk} + c_b
\hat a_s (u)$ for $u < 0$ for some scalar coefficients $c_s$ and
$c_b$.  This is because the matrices involved for $u > (<) 0$ are
then different, so $\hat g$ being continuous at $u=0$ can never be
satisfied. To explain more clearly our idea of solving this
problem, let us first consider the special case $\hat d = \hat z$,
so that $ i (\vec d \cdot \vec \sigma) \sigma_y = \sigma_x$.  Then
$\hat g_{p, \rm bulk}$, $\hat a_p$ above are linear combinations
of $\tau_3, \sigma_x \tau_{\pm}$. To find a possible continuous
$\hat g$ at $u=0$, we must therefore include also decaying
solutions for $u > 0$ which also involves $\sigma_y \tau_{\pm}$
(due to the singlet superconductor on $x < 0$), and exponentially
increasing solution for $u<0$ which also involves $\sigma_x
\tau_{\pm}$. One can find these solutions easily, as done
explicitly in Ref.\cite{Yip93}.  We can however also note that
these needed solutions can be written as $(\sigma_z \tau_3) \hat
a_p$ and $(\sigma_z \tau_3) \hat b_s$. (Note that $ (\sigma_z
\tau_3) $ commutes with $ \tau_3$, $\hat \Delta_s$ and $\hat
\Delta_p$). Hence we can try\cite{Yip93}

\bea
    \hat g(u) &=& \hat g_{s,\rm bulk} + c_s \hat b_s + \zeta_s
    (\sigma_z \tau_3) \hat b_s \qquad u < 0
    \label{an1} \\
    \hat g(u) &=& \hat g_{p,\rm bulk} + c_p \hat a_p + \zeta_p (\sigma_z \tau_3) \hat a_p \qquad u > 0
    \label{an2}
\eea where $c_{s,p}$ and $\zeta_{s,p}$ are scalar coefficients to
be determined.  Note that now $\hat g$ for both $u < (>)0$ consist
of $ \tau_3$, $\sigma_3$, $\sigma_y \tau_{\pm}$ and $\sigma_z
\tau_{\pm}$ matrices and hence a solution is possible.  Note that
eq.\ (\ref{qc2}) is satisfied. Since $\hat g(0)$ can be expressed
as either eq.\ (\ref{an1}) or (\ref{an2}), we can determine the
coefficients $c_{s,p}$ and $\zeta_{s,p}$ using simple algebra, but
a simpler procedure is to left-multiply eq.\ (\ref{an1}) and
(\ref{an2}) (at $u=0$) by $\hat b_s$ and $\hat a_p$ respectively
to obtain

\bea
    \hat b_s \hat g(0) &=& \hat b_s \hat g_{s, \rm bulk} = - i \pi
    \hat b_s
    \label{bg} \\
    \hat a_p \hat g(0) &=& \hat a_p \hat g_{p,\rm bulk} = + i \pi \hat a_p \ .
    \label{ag}
\eea Note that the unknown scalar coefficients have all
disappeared. Further multiplying eq.\ (\ref{bg}) and (\ref{ag})
respectively by $\hat a_s$ and $\hat b_p$, and adding the two
equations, we obtain thus

\be
    \hat g(0) = - i \pi \{ \hat a_p, \hat b_s \}^{-1} [ \hat a_p,
    \hat b_s ] \ . \label{f1}
\ee Repeating the above procedure by post- rather than pre-
multiplication actually shows that we can also reverse the order
of the anticommutator and commutators in eq.\ (\ref{f1}), as can
be also verified explicitly. Note that $\{ \hat a_p, \hat b_s \}$
is a linear combination of $\hat 1$ and $\hat \sigma_z \tau_3$
only.

For $k_x < 0$, $u < 0 (>0)$ corresponds to $x > (< 0)$.  Following again the above
procedure and ensuring that the solutions decay correctly to their respective bulk values
at $u \to \mp \infty$ gives us the analogous formula

\be
    \hat g(0) = - i \pi \{ \hat a_s, \hat b_p \}^{-1} [ \hat a_s,
    \hat b_p ] \ . \label{f2}
\ee Eq.\ (\ref{f1}) and (\ref{f2}) are the special examples of
eq.\ (\ref{g0}) in the present case. For general
$\hat{d}(\hat{k})$, to ensure the  continuity of $\hat{g}$ at
$x=0$, we need matrices $\tau_3$, $\sigma_y \tau_{\pm}$, $(\vec d
\cdot \vec \sigma) \sigma_y \tau_+$ and $\sigma_y (\vec d \cdot
\vec \sigma) \tau_{-}$. A matrix that commutes with $\tau_3$,
$\hat \Delta_s$, $\hat \Delta_p$ can be seen to be
$\left( \ba{cc} (\vec d \cdot \vec \sigma) & 0 \\
0 & \sigma_y (\vec d \cdot \vec \sigma) \sigma_y \ea \right) \equiv
\Sigma_1$. The argument above can be repeated with this matrix
replacing $\sigma_z \tau_3$ above.

The above argument actually does not depend on the fact that
$\sigma_z \tau_3$ (or $\Sigma_1$ defined above) be common to both
sides of eq.\ (\ref{an1}) or (\ref{an2}). To see this, let us
first consider the singlet superconductor.  We note that the
matrices $\hat 1$, $\sigma_y$, $\sigma_x \tau_3$, $\sigma_z
\tau_3$ all commute with $\tau_3$, $\sigma_y \tau_{\pm}$, so they
are automatically solutions to eq.\ (\ref{qcs}).  Since the
product of two solutions to eq.\ (\ref{qcs}) is also a solution,
we see that $\hat g_{s, \rm bulk}$, $\sigma_y \hat g_{s, \rm
bulk}$, $\sigma_x \tau_3 \hat g_{s, \rm bulk}$, $\sigma_z \tau_3
\hat g_{s, \rm bulk}$ are also "constant solutions". There are
also in fact four decaying solutions $\hat a_s$, $\sigma_y \hat
a_s$, $\sigma_x \tau_3 \hat a_s$, $\sigma_z \tau_3 \hat a_s$ and
similarly four exponentially increasing solutions. Note that we
now have $16$ solutions to the $4 \times 4$ matrix equation
(\ref{qcs}), and hence any solution to eq.\ (\ref{qcs}) can be
written in terms of them.  The most general $\hat g$ which decays
to $\hat g_{s, \rm bulk}$ at $u \to - \infty$ (for $k_x > 0$) can
be seen to be

\be
    \hat g(u) = \hat g_{s, \rm bulk} + c_s \hat b_s +
    \zeta_{s,1} \sigma_y \hat b_s + \zeta_{s,2} \sigma_x \tau_3 \hat b_s
    + \zeta_{s,3} \sigma_z \tau_3 \hat b_s
    \label{gan1}
\ee where $c_s$, $\zeta_{s,1-3}$ are scalar coefficients.  Note
that no constant solution other than $\hat g_{s, \rm bulk}$ can
appear on the right hand side of eq.\ (\ref{gan1}) due to the
condition at $u \to - \infty$. Since $\hat 1$, $\sigma_y$,
$\sigma_x \tau_3$, $\sigma_z \tau_3$ all commute with $\tau_3$,
$\sigma_y \tau_{\pm}$, they commute with $\hat b_s$. Left
multiplication of eq.\ (\ref{gan1}) with $\hat b_s$ again yields
eq.\ (\ref{bg}).

The triplet superconductor on $x > 0$ can be treated similarly. For
a given $\hat k$, we have already noted that the matrix
$\left( \ba{cc} (\vec d (\hat k) \cdot \vec \sigma) & 0 \\
0 & \sigma_y (\vec d (\hat k) \cdot \vec \sigma) \sigma_y \ea
\right) \equiv \Sigma_1(\hat k)$ commutes with $\tau_3$, $ (\vec
d(\hat k) \cdot \sigma) \sigma_y \tau_{+}$ and $ \sigma_y ( \vec d
(\hat k) \cdot \sigma) \tau_{-}$. Two other matrices with this
property are (besides $\hat 1$)
$\left( \ba{cc} (\vec d_{2,3} \cdot \vec \sigma) & 0 \\
0 & -\sigma_y (\vec d_{2,3} \cdot \vec \sigma) \sigma_y \ea
\right) \equiv \Sigma_{2,3} (\hat k)$ where $\hat d_{2,3}$ are the
two vectors orthogonal to $\hat d (\hat k)$. These four matrices
$\hat 1$, $\Sigma_{1,2,3}(\hat k)$ are trivial solutions to eq.\
(\ref{qc1}) for the triplet superconductor.  Four other constant
solutions are the product between them and $\hat g_{p, \rm bulk}$.
Again there are four decaying (increasing) solutions obtained by
their product with $\hat a_p$ ($\hat b_p$).  We again have a total
of $16$  solutions to eq.\ (\ref{qc1}) for the triplet
superconductor.  The most general solution to $\hat g(u)$ with
$\hat g(u) \to  \hat g_{p, \rm bulk}$ as $u \to \infty$ (again for
$k_x > 0$) is

\be
    \hat g(u) = \hat g_{p, \rm bulk} + c_p \hat a_p +
    \zeta_{p,1} \Sigma_1(\hat k) \hat a_p + \zeta_{p,2} \Sigma_2 (\hat k) \hat a_p
    + \zeta_{p,3} \Sigma_3 (\hat k) \hat a_p  \ .
    \label{gan2}
\ee On noting that $\Sigma_{1,2,3} (\hat k) $ commute with
$\tau_3$, $ (\vec d(\hat k) \cdot \sigma) \sigma_y \tau_{+}$ and $
\sigma_y ( \vec d (\hat k) \cdot \sigma) \tau_{-}$ and hence $\hat
a_p$, left multiplying eq.\ (\ref{gan2}) by $\hat a_p$ again
yields eq.\ (\ref{ag}). The rest of the demonstration of eq.\
(\ref{f1}) goes through unchanged. Similar argument applies for
$k_x < 0$.

For finite transmission $\mathcal{D}$, $\hat g$ is now in general
discontinuous at $x = 0$, and $\hat g$ for incoming, reflected and
transmitted paths are all related. A general boundary condition
non-linear in $\hat g$ was first derived independently by
Zaitsev\cite{Zaitsev84} and Kieselmann\cite{Kieselmann87}.  In
Ref.\cite{Yip97} a simplified linearized form of the boundary
condition was provided.  The derivation given there was for
singlet superconductors. However, the boundary conditions derived
by Ref.\cite{Zaitsev84,Kieselmann87} were actually independent of
the assumption on the parities of the superconductors. This can
also be checked by using the formulas derived by Millis \emph{et
al.}\cite{Millis88} for a spin-active interface between two
superconductors of different parities.  By ignoring the spin
dependence of the scattering amplitudes in Ref.\cite{Millis88} and
eliminating the "drone amplitudes" there, one can show that the
nonlinear boundary condition of Kieselmann\cite{Kieselmann87} can
be recovered.  This non-linear boundary condition can then be
linearized using arguments similar to those used in
Ref.\cite{Yip97}: We express $\hat g$ in the form  eq.\
(\ref{gan1}) and (\ref{gan2}) for each of the quasiclassical
incident, reflected and transmitted path but with $\hat g_{\rm
bulk}$ replaced by $\hat g_{\rm aux}$, the "auxiliary" solution
corresponding to the completely reflecting case\cite{Yip97} (That
is, for example, $\hat g^r_{\rm aux}$ solves the quasiclassical
equation on the quasiclassical path formed by $\hat {\underline
k}$ and $\hat k$ with the physical order parameter on the right
but with the boundary condition $\hat g^r_{\rm aux} (\hat k) =
\hat g^r_{\rm aux} (\hat {\underline k})$ at $x=0_{+}$). The
decaying and exploding terms can be eliminated using projection
operators\cite{Yip97} with arguments similar to those explained
above for the perfect transmission case. Thus the derivation in
Ref.\cite{Yip97} can be carried over to our present situation. It
is most convenient to write the final results in terms of $\hat
s^{r,l} \equiv \hat g (\hat k, 0_{\pm}) + \hat g (\underline{\hat
k}, 0_{\pm})$ at $x= 0_{\pm}$ and $\hat g_d \equiv \hat g (\hat k,
0_{\pm}) - \hat g (\underline{\hat k}, 0_{\pm})$ where $\hat k$
($\underline{\hat k}$) denotes outgoing reflected (incoming
incident) wavevector on the right ($r$). $\hat k$
($\underline{\hat k}$) is also the incoming (reflected) wavevector
on the left ($l$). $\hat g_d$ (denoted by $\hat d$ in
Ref.\cite{Yip97}) is continuous across the interface and is given
by

\be
    \hat g_d=\frac{\frac{i\mathcal{D}}{2\pi}[\hat{g}^r_{\rm
    aux}, \hat g^l_{\rm aux}]}{1 +
    \frac{\mathcal{D}}{4 \pi^2} (\hat g^r_{\rm aux} - \hat g^l_{\rm
    aux})^2}\:,
    \label{gd}
\ee whereas

\be
    \hat s^r =  \frac{ ( 2 - \mathcal{D})  \hat g^r_{\rm aux} + \mathcal{D} \hat
    g^l_{\rm aux}}
    {1 + \frac{\mathcal{D}}{4 \pi^2}
    (\hat g^r_{\rm aux} - \hat g^l_{\rm aux})^2}\:,\label{sr}
\ee and

\be
    \hat s^l =  \frac{ ( 2 - \mathcal{D})  \hat g^l_{\rm aux} + \mathcal{D} \hat
    g^r_{\rm aux}}
    {1 + \frac{\mathcal{D}}{4 \pi^2}
    (\hat g^r_{\rm aux} - \hat g^l_{\rm aux})^2}\:,\label{sl}
\ee where the subscripts "aux" denote the solution to the
$\mathcal{D}=0$ problem. Since $\hat g_{\rm aux}^{r,l}$ commute
with the anti-commutator $\{\hat g^r_{\rm aux} , \hat g^l_{\rm
aux} \} $, we need not specify the relative order between the
numerator and the denominator in eqs.\ (\ref{gd})-(\ref{sl}).

By some straightforward algebra, the complete quasiclassical
Green's function $\hat{g}^r_{\rm{aux}}$ for $\mathcal{D}=0$
problem in Sec.\ \ref{noD} is shown to be,

\bea
    \hat{g}^r_{\rm{aux}}&=&\frac{(-i\pi)}{\epsilon_n^2+|\Delta_p|^2\sin^2\phi}
    [\alpha_p\epsilon_n\tau_3+\frac{i}{2}|\Delta_p|^2\sin(2\phi)\sigma_z\tau_3\\\nonumber
    &+&{\Delta_p}(\alpha_p\sin\phi\ \sigma_x+\epsilon_n\cos\phi\
    \sigma_y)\sigma_y\tau_+
    +{\Delta_p^*}\sigma_y
    (\alpha_p\sin\phi\ \sigma_x-\epsilon_n\cos\phi\ \sigma_y
    )\tau_{-}
    ]\:,
\eea which can be shown to satisfy
$(\hat{g}^r_{\rm{aux}})^2=-\pi^2$. Together with the trivial
$\hat{g}^l_{\rm{aux}}=\frac{-i\pi}{\alpha_s}(\epsilon_n\tau_3-\Delta_s\sigma_y\tau_+-\Delta_s^*\sigma_y\tau_-)$
for the left side, one can obtain $\hat{g}_d$ in the following
form,

\be
    \hat{g}_d=(-i\pi\mathcal{D})\hat{C}\hat{A}\:,
\ee where the matrix $\hat{C}$ is the inverse of

\be
    \hat{C}^{-1}\equiv(2-\mathcal{D})\alpha_s
    \left[\epsilon_n^2+(|\Delta_p|\sin\phi)^2\right]
    +\frac{\mathcal{D}}{2}\hat{B}\:.
\ee The matrix $\hat{B}$ comes from the anticommutator and is
given by,

\bea
    \hat{B}&\equiv&\alpha_s\left[\epsilon_n^2+(|\Delta_p|\sin\phi)^2\right]
    \frac{\{\hat{g}^r_{\rm{aux}},\hat{g}^l_{\rm{aux}}\}}{(-\pi^2)}\\\nonumber
    &=&
    2\alpha_p\epsilon_n^2+i\epsilon_n|\Delta_p|^2\sin(2\phi)\
    \sigma_z
    -2i\epsilon_n|\Delta_s||\Delta_p|\cos\phi\sin\chi\
    \sigma_y\\\nonumber
    &-&2\alpha_p|\Delta_s||\Delta_p|\sin\phi\cos\chi\ \sigma_x\tau_3
    +|\Delta_p|^2\sin(2\phi)(\Delta_s^*\ \sigma_x\tau_--\Delta_s\
    \sigma_x\tau_+)\:.
\eea $\hat{A}$ is from the following commutator,

\bea
    \hat{A}&\equiv&\alpha_s\left[\epsilon_n^2+(|\Delta_p|\sin\phi)^2\right]
    \frac{\left[\hat{g}^r_{\rm{aux}},\hat{g}^l_{\rm{aux}}\right]}{(-\pi^2)}\\\nonumber
    &=&
    -2\epsilon_n|\Delta_s||\Delta_p|\cos\phi\cos\chi\
    \sigma_y\tau_3
    -2i\alpha_p|\Delta_s||\Delta_p|\sin\phi\sin\chi\
    \sigma_x\\\nonumber
    &-&
    2\left[\epsilon_n\Delta_p(\alpha_p\sin\phi\ \sigma_x+\epsilon_n\cos\phi\ \sigma_y)
    +\alpha_p\epsilon_n\Delta_s\right]\sigma_y\tau_+\\\nonumber
    &-&
    2\sigma_y\left[\epsilon_n\Delta_p^*(-\alpha_p\sin\phi\ \sigma_x+\epsilon_n\cos\phi\ \sigma_y)
    -\alpha_p\epsilon_n\Delta_s^*\right]\tau_-\:.
\eea Similarly, the sum can be expressed as,

\be
    \hat{s}^r=4\hat{C}\left[
    (1-\frac{\mathcal{D}}{2})\hat{g}^r_{\rm{aux}}
    +\frac{\mathcal{D}}{2}\hat{g}^l_{\rm{aux}}
    \right]\alpha_s(\epsilon_n^2+|\Delta_p|^2\sin^2\phi)\:,
\ee and the expression for $\hat{s}^l$ is identical to the above
with interchange of $(1-\frac{\mathcal{D}}{2})$ and
$\frac{\mathcal{D}}{2}$ in the bracket.

%In eq.\ (\ref{current})-(\ref{spincurrent}), the trace contributions
%from the particle and hole sectors are supposed to be equal, and
%hence we can consider, alternatively, the trace within the particle
%sector only. For the supercurrent $J_x$, namely, the part of trace
%in eq.\ (\ref{current}) can be rearranged as
%${\rm{Tr}}\left[\frac{1+\tau_3}{2}(\hat{g}(\hat{k},\epsilon_n)-\hat{g}(-\hat{k},-\epsilon_n))\right]$,

%\be
%    J^i_j\sim
%    \sum_{\hat{k}}T\sum_{\epsilon_n}\hat{k}_j{\rm{Tr}}\{\sigma_i\frac{1+\tau_3}{2}
%    \left[\hat{g}(\hat{k},\epsilon_n)-\hat{g}(-\hat{k},-\epsilon_n)\right]\}
%\ee and using the relation $\hat{g}(-\hat{k},
%-\epsilon_n)=\tau_2\hat g^{tr}(\hat{k},\epsilon_n)\tau_2$, we find
%that $J^x_y$ can exist only when the $\sigma_1\tau_3$ component
%appears in $\hat{g}$. For $J^y_x$, it takes the $\sigma_2$
%component.

\newpage

%\newpage

%\newpage

\begin{thebibliography}{plain}

\bibitem{review} See the review articles, Y. K. Kato and D. D. Awschalom, J. Phys. Soc. Jpn. {\bf{77}},
031006 (2008); M. K\"onig, H. Buhmann, W. Molenkamp, T. Hughes, C.
X. Liu, X.-L. Qi, and S. C. Zhang, {\it ibid} {\bf{77}}, 031007;
N. Nagaosa, {\it ibid} {\bf{77}}, 031010.

%\bibitem{Vollhardt} Vollhardt and W\"olfle, Helium Three

\bibitem{group}  E. I. Blount, Phys. Rev. B {\bf 32}, 2935 (1985);
G. Volovik and L. P. Gorkov, Zh. Eksp. Teor. Fiz. {\bf 88}, 1412
(1985) [Sov. JETP {\bf 61}, 843 (1985)]; S. Yip and A. Garg, Phys.
Rev. B {\bf 48}, 3304 (1993)

\bibitem{Bauer-r} For a review on CePt$_3$Si, see E. Bauer, H. Kaldarar, A. Prokofiev, E.
Royanian, A. Amato, J. Sereni, W. Br\"{a}mer-Escamilla, and I.
Bonalde, J. Phys. Soc. Jpn. {\bf{76}}, 051009 (2007).

\bibitem{Ce113} R. Settai, T. Takeuchi, and Y. \={O}nuki, J. Phys.
Soc. Jpn. {\bf{76}}, 051003 (2007)

\bibitem{Reyren} N. Reyren, S. Thiel, A. D. Caviglia, L. Fitting Kourkoutis,
G. Hammerl, C. Richter, C. W. Schneider, T. Kopp, A. -S.
R\"{u}etschi, D. Jaccard, M. Gabay, D. A. Muller, J. -M. Triscone,
J. Mannhart, Science {\bf 317}, 1196 (2007).

\bibitem{Vorontsov} A. B. Vorontsov, I. Vekhter, and M. Eschrig, Phys. Rev. Lett.
{\bf{101}}, 127003 (2008).

\bibitem{Tanaka} Y. Tanaka, T. Yokoyama, A. V. Balatsky, and N. Nagaosa,
Phys. Rev. B {\bf{79}}, 060505(R) (2009).

\bibitem{e-m} V. M. Edelstein, Sov. Phys. JETP {\bf 68}, 1244 (1989),
Phys. Rev. Lett. {\bf 75}, 2004 (1995);  S. K. Yip, Phys. Rev. B {\bf 65},
 144508 (2002).

\bibitem{QSH} C. L. Kane and E. J. Mele, Phys. Rev.  Lett.
 {\bf{95}}, 146802 (2005); {\it{ibid}} {\bf{95}}, 226801 (2005).

\bibitem{QSH2} S. Murakami, S. Iso, Y. Avishai, M. Onoda, and N. Nagaosa,
Phys. Rev. B {\bf{76}}, 205304 (2007).

\bibitem{helicalSC} M. Stone and R. Roy, Phys. Rev. B {\bf{69}},
184511 (2004); X.-L. Qi, T. L. Hughes, S. Raghu, and S. C. Zhang,
arXiv:0803.3614

%\bibitem{Qi} X. L. Qi, T. L. Hughes, S. Raghu, and S. C. Zhang, arXiv:0803.3614

\bibitem{Sato} M. Sato and S. Fujimoto, Phys. Rev. B {\bf{79}},
094504 (2009).

\bibitem{Sengupta} K. Sengupta and V. M. Yakovenko, Phys. Rev.
Lett. {\bf{101}}, 187003 (2008).

\bibitem{Yip93} S.-K. Yip, J. Low Temp. Phys. {\bf 91}, 203 (1993)

\bibitem{Yip90} S.-K. Yip, O. F. De Alcantara Bonfim and P. Kumar,
  Phys. Rev. B {\bf 41}, 11214 (1990)

\bibitem{Serene} J. W. Serene and D. Rainer, Phys. Rep.  {\bf 101}, 221  (1983)

\bibitem{unequal} L.-F. Chang and P. F. Bagwell, Phys. Rev. B {\bf{49}}, 15853 (1994);
S. K. Yip, Phys. Rev. B {\bf{68}}, 024511 (2003); S.-T. Wu and S. Yip, Phys. Rev.
B {\bf{70}}, 104511 (2004).

\bibitem{Thuneberg} E. V. Thuneberg, J. Kurkij\"arvi and D. Rainer,
 Phys. Rev. B {\bf 29}, 3913 (1984)

\bibitem{Fogelstrom} M. Fogelstr\"om and J. Kurkij\"arvi, J. Low Temp. Phys. {\bf 98}, 195 (1995)


\bibitem{Yip97} S.-K. Yip, J. Low Temp. Phys. {\bf 109}, 547 (1997)

\bibitem{footnote} Physical quantities can be obtained by using only
the particle-particle block of $\hat g$, but these forms would be
more convenient below.  In deriving these two relations eq.\
(\ref{magnetization}) and (\ref{spincurrent}), we have used the
symmetry\cite{Serene} $\hat{g}(-\hat{k}, -\epsilon_n)=\tau_2\hat
g^{tr}(\hat{k},\epsilon_n)\tau_2$.

\bibitem{Zaitsev84}  A. V. Zaitsev, {\it Zh. Eksp. Teor. Fiz.} {\bf
86}, 1742 (1984) [ Sov. Phys. JETP, {\bf 59}, 1015 (1984)].

\bibitem{Kieselmann87} G. Kieselmann, Phys. Rev. B {\bf 35}, 6762
(1987); Ph. D. Thesis, U. Bayreuth, unpublished.

\bibitem{Millis88} A. Millis, D. Rainer and J. A. Sauls, Phys.
Rev. B {\bf 38}, 4504 (1988)


\end{thebibliography}
\end{document}